\documentclass{aa}
\usepackage{graphicx}
\usepackage{natbib}
\usepackage{amssymb,amsmath}
\usepackage{color}
\bibpunct{(}{)}{;}{a}{}{,}

\begin{document}

\title{Effects of anisotropic winds on massive stars evolution}
\titlerunning{Effects of anisotropic winds on massive stars evolution}

\author{Cyril Georgy \and Georges Meynet \and Andr\'e Maeder}
\authorrunning{Georgy et al.}

\institute{Geneva Observatory, Geneva University, CH--1290 Sauverny, Switzerland\\
              email: Cyril.Georgy@unige.ch\\
               }

\date{Received  / Accepted }

\offprints{Cyril Georgy} 
\abstract{Whenever stars are rotating very fast ($\Omega/\Omega_\mathrm{crit} > 0.7$, with $\Omega_\mathrm{crit}$ the Keplerian angular velocity of the star accounting for its deformation) radiative stellar winds are enhanced in polar regions. This theoretical prediction is now confirmed by interferometric observations of fast rotating stars.}
{Polar winds remove less angular momentum than spherical winds and thus allow the star to keep more angular momentum. We quantitatively assess the importance of this effect.}
{First we use a semi-analytical approach to estimate the variation of the angular momentum loss when the rotation parameter increases. Then we compute complete 9 M$_\odot$ stellar models at very high angular velocities (starting on the ZAMS with $\Omega/\Omega_\mathrm{crit} = 0.8$ and reaching the critical velocity during the Main Sequence) with and without radiative wind anisotropies.}
{When wind anisotropies are accounted for, the angular momentum loss rate is reduced by less than $4\%$ for $\Omega/\Omega_\mathrm{crit} < 0.9$ with respect to the case of spherical winds. The reduction amounts to at most  $30\%$ when the star is rotating near the critical velocity. These values result from two counteracting effects:  on the one hand polar winds reduce the loss of angular momentum, on the other hand, surface deformations imply that the mass which is lost at high co-latitude is lost at a larger distance from the rotational axis and thus removes more angular momentum.}
{In contrast with previous studies, which neglected surface deformations, we show that the radiative wind anisotropies have a relatively modest effect on the evolution of the angular momentum content of fast rotating stars.}


\maketitle

\section{Introduction}

Over the last years, we added to the Geneva stellar evolution code several major improvements, as rotation, the inclusion of extended nuclear reaction network allowing to explore the advanced phases of massive star evolution \citep[neon, oxygen and silicon burning, see][]{Hirschi2004a}, and the inclusion of magnetic field . 

The inclusion of rotation improved the agreement between the outputs of numerical models and  observational results, as the surface enrichments, the ratio of blue to red supergiants in the SMC, the ratio of WR to O-type stars and the ratio of type Ibc to type II supernovae \citep[see, for example][]{Meynet2000a,Maeder2001a,Meynet2003a,Georgy2009a}. The treatment of internal magnetic field in the models gives a more realistic rate of Gamma Ray Bursts as a function of the metallicity \citep{Yoon2006a}, produces a rotation rate for the young pulsars in much better agreement with the observations \citep{Heger2005a}, and allows to explain the flat rotation profile of the Sun \citep{Eggenberger2005a}.

Rotation acts not only in the interior of the stars, however, but has also several effects on the surface. \citet{Maeder2000a} show that rotation increases the global mass loss rate. It also modifies the shape of the star, and consequently surface quantities such as: effective gravity, effective temperature, and radiative flux. Interestingly, \citet{Maeder2002a} shows that the mass loss in fast rotating massive stars does not remain isotropic, but becomes increasingly anisotropic as the rotation approaches the critical rotation parameter. They show that this favours a bipolar stellar wind, and modifies the quantity of angular momentum removed from the star.

With the developments of the interferometric technics, some of the predicted effects have recently become observable. For example, \citet{Carciofi2008a} have obtained for the ratio of the equatorial to polar radius a value of $1.5$ for the very fast rotating star Achernar as it is expected in the frame of the Roche model theory. \citet{Monnier2007a} provide a map of the effective temperature over the surface of Altair, showing that the temperature gradient between the pole and the equator of this star is in good agreement with the von Zeipel theorem \citep{vonZeipel1924a}. \citet{Meilland2007a} observed evidences of a disc and of a polar wind around the star $\mu\,\mathrm{Ara}$, as is predicted for a star rotating at the critical limit.

At the moment the effects of wind anisotropies on stellar models have been quantitatively explored only in two previous publications \citep{Meynet2003a,Meynet2007a}. In the present work we improve the numerical treatment over these works and reexamine this effect. We include the following improvements:
\begin{itemize}
\item For the first time, we account not only for the variation of the mass flux with the co-latitude as was done in the previous work, but we also account for the surface deformation of the star. As we will see this last effect cannot be neglected.
\item We use an updated expression for the mass flux obtained by \citet{Maeder2009a}.
\item We accounted for the variation of the force multiplier parameters over the surface of fast rotating stars (see below for more details on that point).
\end{itemize}

To check the validity of the numerical results and also to study in a clearer and simpler context the various effects intervening in the loss of angular momentum, we propose a semi-analytical approach to estimate the effect of fast rotation on the loss of angular momentum. The relative effects obtained in that way depend only on one parameter, the ratio $\Omega/\Omega_{\rm crit }$, where $\Omega$ is the surface angular velocity and $\Omega_{\rm crit }$, the critical angular velocity, \textit{i.e.}, the angular velocity such that the centrifugal acceleration at the equator compensates for the gravity (accounting for the deformation of the shape of the star).

The paper is organised in the following way: in Sect.~\ref{theory}, we give the theoretical aspects of wind anisotropy. The third section presents our semi-analytical approach. In Sect.~\ref{evolution}, we discuss results based on complete numerical stellar models. Conclusions are presented in Sect.~\ref{conclusion}.

\section{Rotation and wind anisotropy}\label{theory}

\subsection{Increase of the global mass loss rate induced by rotation}

As shown by \citet{Maeder2000a}, the local radiative mass loss rate $\Delta\dot{M}$ by unit surface $\Delta\sigma$ can be written:
\begin{equation}
\frac{\Delta\dot{M}}{\Delta\sigma}\sim A\left(\frac{ac}{4}\right)^\frac{1}{8}\left[\frac{L}{4\pi GM_\star}\right]^{\frac{1}{\alpha}-\frac{1}{8}}\frac{g_\mathrm{eff}^{1-\frac{1}{8}}\left[1+\zeta(\theta)\right]^\frac{1}{\alpha}}{\left(1-\Gamma_\Omega(\theta)\right)^{\frac{1}{\alpha}-1}}.\label{Mdotsurf}
\end{equation}
In this expression $A = (k\alpha)^\frac{1}{\alpha}\left(\frac{1-\alpha}{\alpha}\right)^\frac{1-\alpha}{\alpha}$, where $\alpha$ and $k$ are the force multiplier parameters empirically determined \citep{Lamers1995a}, $M_\star = M\left(1-\frac{\Omega^2}{2\pi G\rho_\mathrm{m}}\right)$ is the reduced mass, where $\rho_\mathrm{m}$ is the internal average density, $L$ is the stellar luminosity, $g_\mathrm{eff}$ is the effective gravity at the co-latitude $\theta$ (\textit{i.e.}, the vectorial sum of the gravitational acceleration and of the centrifugal one) and $\zeta(\theta)$ expresses the deviation from the von Zeipel theorem produced by shellular rotation \citep[][this term is generally negligible]{Maeder1999a},  $\Gamma_\Omega(\theta)$ is the local Eddington factor, taking into account the effect of rotation:
 
\begin{equation}
\Gamma_\Omega(\theta) = \frac{\kappa_\mathrm{es}L}{4\pi cGM\left(1-\frac{\Omega^2}{2\pi G\rho_\mathrm{m}}\right)},\label{GammaOmega}
\end{equation}
with $\kappa_\mathrm{es}$ the electron scattering opacity.The term $1/8$ in the power of expression (\ref{Mdotsurf}) was added by \citet[][see his chapter 14.4]{Maeder2009a} and does not appear in \citet[][a $T_\mathrm{eff}^{-\frac{1}{2}}$ was absent in their expression 4.24]{Maeder2000a}. Note that in the frame of the line driven wind theory used here to obtain the expression of the mass flux (Eq. (\ref{Mdotsurf})), the opacity is expressed in as a function of the electron scattering opacity \citep{Castor1975a}. The variations are accounted for through the force multiplier parameters $\alpha$ and $k$.

Averaging expression (\ref{Mdotsurf}) over the whole stellar surface $\Sigma$, one obtains the total mass loss rate of the star 
\begin{equation}
\dot{M}\sim\frac{AL^{\frac{1}{\alpha}-\frac{1}{8}}\Sigma^\frac{1}{8}}{\left(4\pi GM\left[1-\frac{\Omega^2}{2\pi G\rho_\mathrm{m}}\right]\right)^{\frac{1}{\alpha}-1}\left(1-\Gamma_\Omega\right)^{\frac{1}{\alpha}-1}}.
\end{equation}
This allows us to compute the ratio of the mass loss rate of a rotating star to the mass loss rate of a non-rotating one lying at the same position in the Hertzsprung--Russel diagram:
\begin{equation}
\frac{\dot{M}(\Omega)}{\dot{M}(\Omega=0)} = \frac{\left(1-\Gamma_\mathrm{Edd}\right)^{\frac{1}{\alpha}-1}}{\left(1-\frac{\Omega^2}{2\pi G\rho_\mathrm{m}}\right)^{\frac{1}{\alpha}-\frac{7}{8}}\left(1-\Gamma_\Omega\right)^{\frac{1}{\alpha}-1}},\label{MM0}
\end{equation}
where $\Gamma_\mathrm{Edd}$ is the classical Eddington factor for a non-rotating star. As a result, we see that the faster the star rotates, the more mass will be lost per time unit.

\subsection{Wind anisotropy}\label{Aniso}

According to \citet{Maeder1999a}, two main effects contribute to the development of anisotropies in the stellar winds. The first, called the $g_\mathrm{eff}$-effect, is due to the variation of the effective gravity with the co-latitude: $g_\mathrm{eff}$ is smaller at the equator than at the poles, and thus, the mass loss, which is directly related to $g_\mathrm{eff}$ (see eq.~(\ref{Mdotsurf})), is favoured at the poles for a rotating star.

The second effect is called the $\kappa$-effect. Due to the so-called bistability in the stellar winds \citep[see][]{Lamers1995a}, the $A$ term in eq.~(\ref{Mdotsurf}) increases for lower values of the effective temperature, \textit{i.e.}, towards the equatorial regions \citep[see Fig.~6 in][]{Ekstrom2008b}. This occurs when the effective temperature are below $11500\,\mathrm{K}$ (Lamers, private com.). This favours an equatorial mass loss.

Looking at eq.~(\ref{Mdotsurf}), we also expect a contribution to the latitudinal variation of the mass loss due to the term $\kappa(\theta)$ in $\Gamma_\Omega(\theta)$.

\subsection{Critical velocities}

As the concept of critical velocity is treated in very different ways through the literature, we briefly repeat some common definitions. Following \citet{Maeder2000a}, we define two critical velocities. The first one is the traditional Keplerian velocity at the equator when the star rotates at the critical velocity defined by $g_\mathrm{eff} = 0$:
\begin{equation}
v_\mathrm{crit,1}=\sqrt{\frac{GM}{R_\mathrm{eb}}}=\sqrt{\frac{2GM}{3R_\mathrm{pb}}},\label{vcrit1}
\end{equation} 
where $R_\mathrm{eb}$ ($R_\mathrm{pb}$) is the equatorial (polar) radius when the first critical velocity is reached. The numerical factors $2/3$ comes from the polar to equatorial radius ratio when the star is at the critical velocity and the Roche approximation is valid \citep[see, e.g.,][]{Ekstrom2008b}. We also define the critical angular velocity $\Omega_\mathrm{crit} = \frac{v_\mathrm{crit,1}}{R_\mathrm{eb}}$, and the ratio $\omega = \frac{\Omega}{\Omega_\mathrm{crit}}$.

The second critical velocity is reached when the star is at the so-called $\Omega\Gamma$--limit, \textit{i.e.} when the local Eddington factor (accounting for the effects of rotation) defined in eq.~(\ref{GammaOmega}) is 1. According to \citet{Maeder2000a}, this term is  equal to:
\begin{equation}
v_\mathrm{crit,2}^2 = \frac{81}{16}\frac{1-\Gamma_\mathrm{Edd}}{V_\mathrm{b}}\frac{GM}{R_\mathrm{eb}^3}R_\mathrm{e}^2,\label{vcrit2}
\end{equation}
with $\Gamma_\mathrm{Edd}$ the Eddington factor, $V_\mathrm{b}$ is the ratio of the volume enclosed by the surface when the star rotates at the second critical velocity to the volume of a sphere with a radius equal to $R_\mathrm{pb}$ , $R_\mathrm{e}$ is the actual equatorial radius. For $\Gamma_\mathrm{edd} \le 0.639$, $v_\mathrm{crit,2} = v_\mathrm{crit,1}$. For $\Gamma_\mathrm{Edd} > 0.639$, $v_\mathrm{crit,2} < v_\mathrm{crit,1}$, and the relevant critical velocity is the second one, because it is reached first.

\section{Effects of fast rotation on angular momentum loss: a semi-analytical approach}

In this section we derive the variations of 
\begin{itemize}
\item the shape of the star,
\item the radiative mass flux,
\item the angular momentum flux,
\item the global angular momentum loss,
\end{itemize}
as a function of only one parameter: $\omega$. Said in other words, the results obtained are independent of the mass, metallicity, and evolutionary stage of the star considered. To obtain such a simple dependence, some hypothesis and normalizations have to be made:
\begin{itemize}
\item We use the Roche approximation for computing the gravitational potential \citep[which is valid here, see][]{Meynet2010b}. Note that we also assume the Roche approximation in our numerical stellar models. This assumption is not only supported by the direct observations of (a few) rapidly rotating stars, but also by the fact that, in our numerical models, the angular velocities are far below the critical velocity in a large fraction of the total mass. The results presented in this work are however probably dependent on the validity of this hypothesis.
\item We neglect the variations over the surface of the force multiplier parameters (see below). This variation will be accounted for in more complex stellar evolution models (see Section~\ref{MdotVink}).
\item We neglect the correcting factor $\zeta(\theta)$ in Eq. \ref{Mdotsurf}.
\item The angular velocity of the surface is assumed to not depend on the colatitude $\theta$ (no differential rotation of the surface).
\item We suppose that the polar radius of the star remains  constant when $\omega$ increases from $0$ (no rotation) to $1$ (critical rotation). This is well verified in complete numerical models \citep{Ekstrom2008b}. We normalise the polar radius $R_\mathrm{p}$ to $1$.
\item The total mass loss rate is taken equal to $4\pi$ for all rotation velocities. In that case, the mass loss rate per unit surface is equal to $1$ in the non-rotating case.
\end{itemize}

\subsection{Shape of the surface}

\begin{figure}[t]
\begin{center}
\includegraphics[width=9cm]{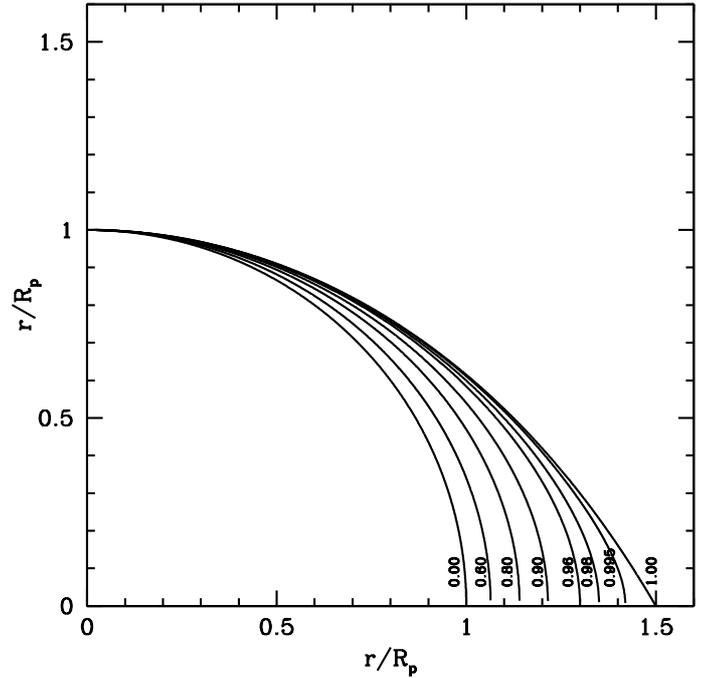}
\caption{Shape of the surface for various values of $\omega = \frac{\Omega}{\Omega_\mathrm{crit}}$ (labelled at the bottom of each curve). The x-axis is the equatorial radius, and the y-axis the polar one. Hence this is how we would see the star equator-on.}
\label{FigSurf}
\end{center}
\end{figure}

\begin{figure*}[t]
\begin{center}
\includegraphics[width=.47\textwidth]{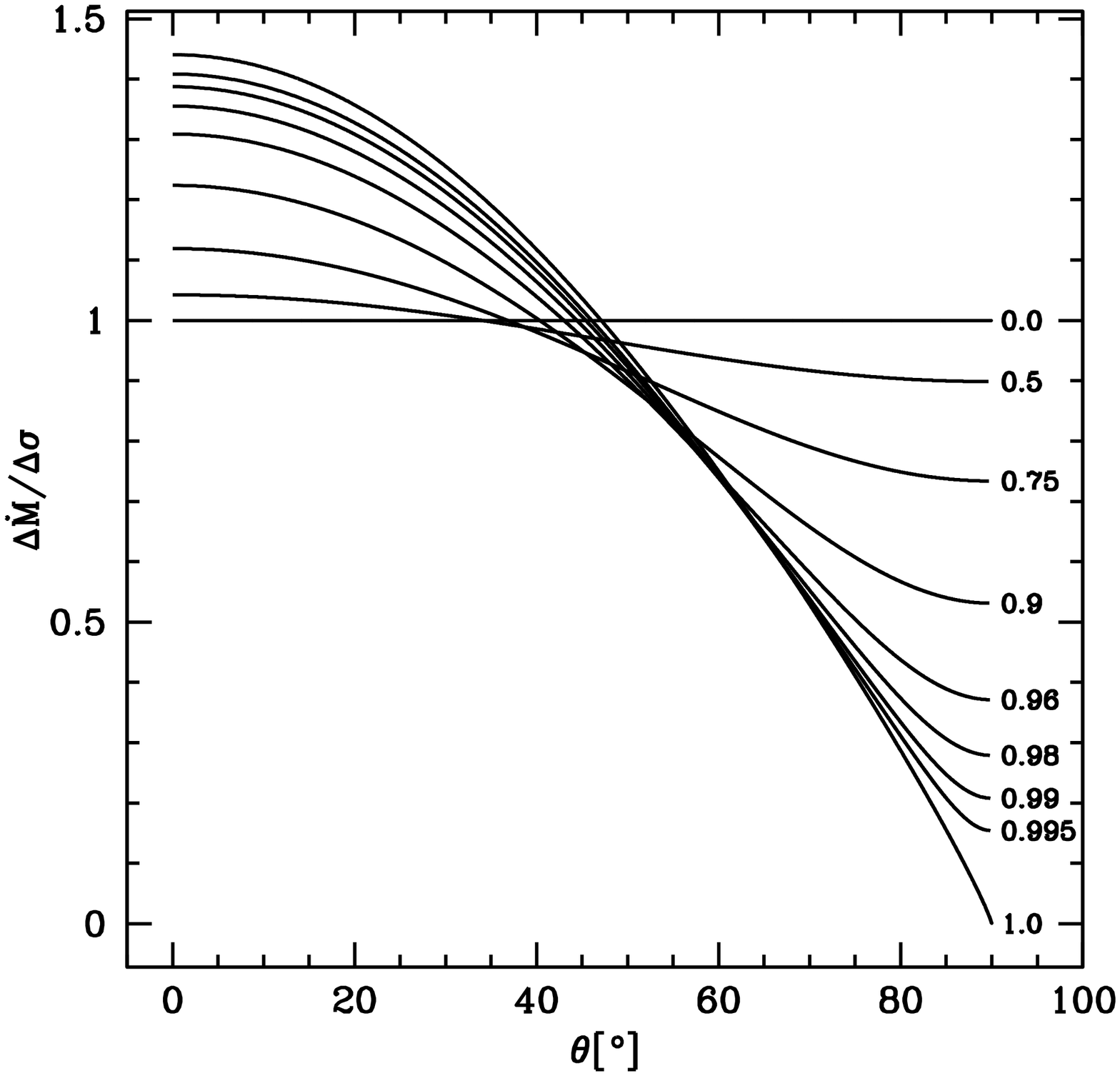}\includegraphics[width=.53\textwidth]{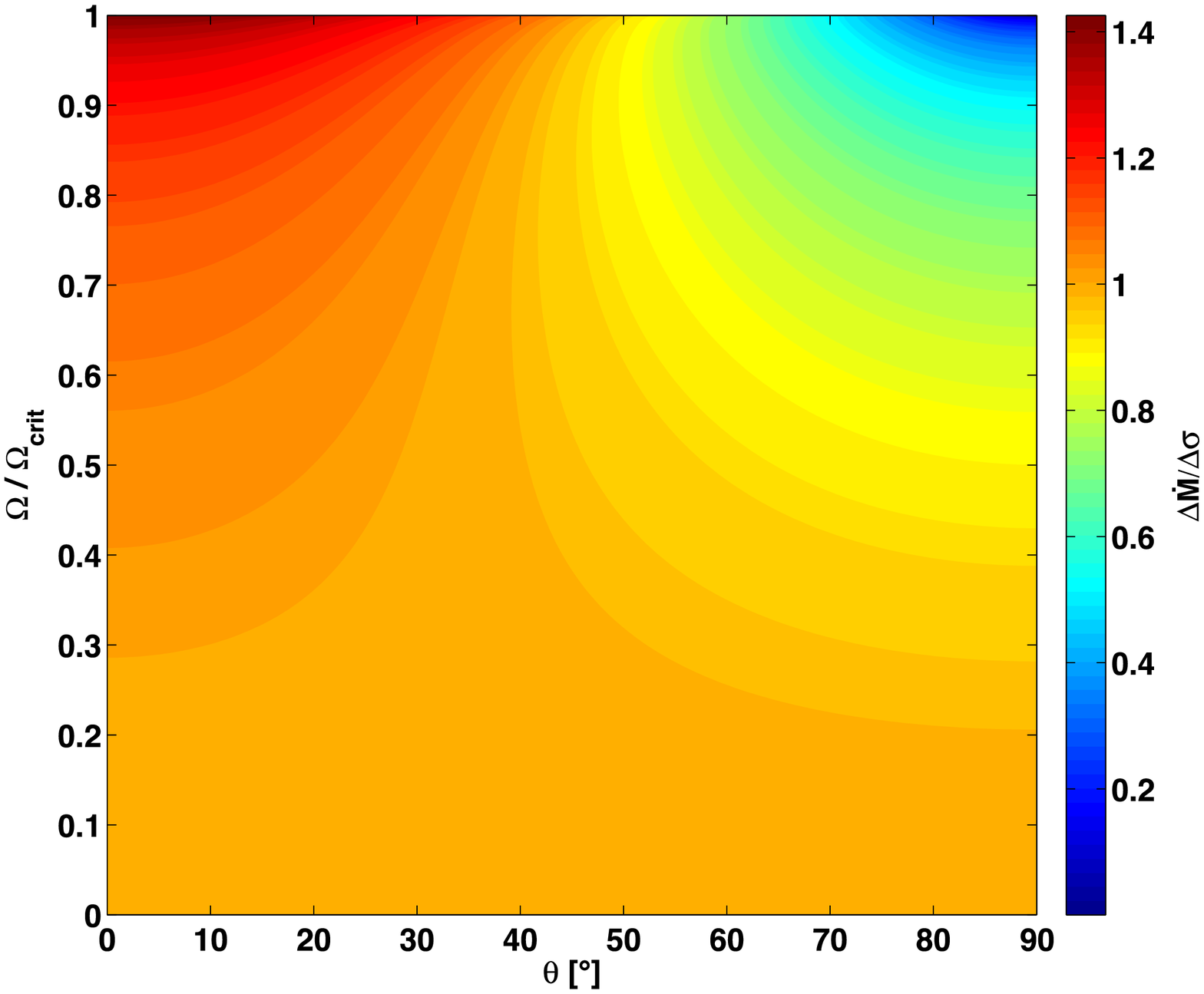}
\caption{\textit{Left panel:} Mass loss rate per surface unit as a function of the co-latitude $\theta$. The pole is on the left, the equator on the right. The mass loss rate per surface unit is normalised to $1$ in the non-rotating case, and thus is expressed without unit. \textit{Right panel: } 2D colour plot of the mass loss rate per surface unit. The $x$-axis is the co-latitude, the $y$-axis the rotation parameter $\omega=\Omega/\Omega_\mathrm{crit}$. The colour scale on the right indicates the local mass loss rate per surface unit: from blue for the lower mass loss flux to red for the higher.}
\label{MdotFig}
\end{center}
\end{figure*}

With the above hypothesis, the equation of the surface of the star can be given as a function of the rotation rate $\omega$ \citep{Maeder2002a}
\begin{equation}
\frac{1}{x(\omega,\theta)} + \frac{4}{27}\omega^2x^2(\omega,\theta)\sin^2(\theta) = 1,\label{Surf}
\end{equation}
where $x(\omega,\theta) = \frac{r(\omega,\theta)}{R_\mathrm{p}}$ is the ratio of the radius at a given co-latitude to the polar one. We can easily express $\theta$ as a function of the normalised radius $x$:
\begin{align}
\theta(x) &= \arcsin\left(\sqrt{\frac{27(x-1)}{4\omega^2x^3}}\right)\quad&\text{if }\omega\not= 0\notag\\
x(\theta) &= 1 &\text{if }\omega = 0.\label{Shape}
\end{align}
The range of satisfactory values for $x$ is a function of $\omega$. It starts from $1$ (to have a positive value under the square root), and goes up to the first positive root of the equation $4\omega^2 x^3 -27x + 27 = 0$. In Fig.~\ref{FigSurf}, we show how the shape of the surface varies for various values of $\omega$, starting from $\omega = 0$ (non rotating case) to $\omega=1$ (critically rotating case). As $\omega$ increases, the centrifugal force deforms ever more the star, and the equatorial radius increases. When the star is exactly at the critical angular velocity ($\omega=1$), we see from eq.~(\ref{Surf}) that the equatorial radius is $1.5$ times larger than the polar one.
 
Eq.~\ref{Surf}, which allows to deduce the shape of the star, depends only on $\omega$. Maximal deformation is obtained for $\omega=1$. Stars reaching the second critical velocity  will have a ratio $\omega$ below $1$. Despite being at the critical velocity, they will not show as strong deformations as stars reaching the first classical critical limit. Since, as we will see, it is the deformation of the star which triggers the wind anisotropies. This means that stars which would be at the $\Omega\Gamma$--limit do not present as strong wind anisotropies as stars at the classical $\Omega$-limit. Accordingly, the fact that $\eta$-Carinae presents strong polar winds, implies that this star should rotate at velocities close to the first classical critical rotation velocity, in case the bipolar shape is due to rotationally induced wind anisotropy.

\subsection{Mass flux variations with the latitude}

\begin{figure}[h]
\begin{center}
\includegraphics[width=9cm]{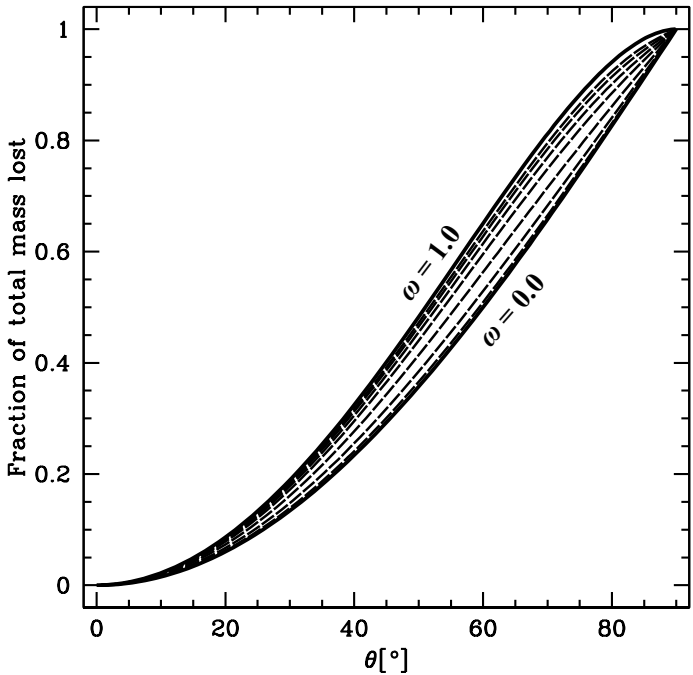}
\caption{Fraction of the total mass loss contained in a cone with a semi-aperture $\theta$ given by the $x$-axis. The lower curve is the result for a non-rotating star losing its mass isotropically. The upper curve is for a critically rotating star ($\omega=1$). The intermediate dashed curves are the results of the same rotation parameter omega as in Fig.~\ref{MdotFig}, \textit{i.e.} $\omega=0.5,\,0.75,\,0.9,\,0.96,\,0.98,\,0.99,\,0.995$ respectively (from bottom to top).}
\label{MassIn}
\end{center}
\end{figure}

To study how the mass flux is modified by the rotation, we use eq.~\ref{Mdotsurf} to compute the local mass loss rate per unit surface. As said above, we normalise the total mass loss to a value of $4\pi$, in order to have a local mass flux of $1$ at every co-latitude in the non-rotating case.

The results of these calculations are shown in Fig.~\ref{MdotFig}. Not surprisingly, the mass loss rate per surface unit is constant when there is no rotation. As the rotation parameter increases, the mass flux increases towards the pole, and decreases towards the equator, due to the variation of the effective gravity, producing a strong anisotropy in the winds. Typically the mass flux at the pole is greater than two times the mass flux at the equator for $\omega \gtrsim 0.8$. In the extreme case, when the rotation parameter $\omega=1$, the effective gravity at the equator is zero, and the radiative mass flux becomes also zero in this idealised representation.

From Fig.~\ref{MdotFig}, we see that the faster the star rotates, the more mass will be lost in the polar region. Integrating the mass flux from the pole to a given co-latitude $\theta$ and dividing by the total mass loss rate, we obtain the fraction of the total mass flux in a cone of semi-aperture $\theta$, for a given rotation factor $\omega$:
\begin{equation}
\frac{\dot{M}_\mathrm{0-\theta}}{\dot{M}_\mathrm{tot}} = \frac{\int_0^\theta\frac{\Delta\dot{M}}{\Delta\sigma}(\vartheta)\mathrm{d}\sigma(\vartheta)}{\dot{M}_\mathrm{tot}},
\end{equation}
where $\mathrm{d}\sigma(\theta)$ is the surface element at the colatitude $\theta$, given by $\mathrm{d}\sigma = \frac{r^2(\theta)\sin(\theta)\mathrm{d}\varphi\mathrm{d}\theta}{\cos(\varepsilon)}$, with $\varepsilon$ the angle between the local effective gravity and the radial direction. The angle $\varepsilon$ is computed using the components of the effective gravity $g_\mathrm{eff}$:
\begin{equation}
\cos(\epsilon) = \frac{\vec{g}_\mathrm{eff} \cdot \vec{e}_r}{||\vec{g}_\mathrm{eff}||}.
\end{equation}
with
\begin{align}
\vec{g}_\mathrm{eff} =& \left(-\frac{GM}{r^2(\theta)}+\Omega^2 r(\theta)\sin^2(\theta)\right)\vec{e}_r\notag\\
&+ \Omega^2 r(\theta)\sin(\theta)\cos(\theta)\vec{e}_\theta,
\end{align}
where $\vec{e}_r$ ($\vec{e}_\theta$) is the radial (colatitudinal) unit vector.

The result is shown in Fig.~\ref{MassIn}. For a non--rotating star (lower curve), we see that $50\%$ of the total mass is lost in a cone of semi-aperture $60^\circ$. When the star is at the first critical velocity, the aperture of the cone containing $50\%$ of the total mass flux is slightly reduced: it is around $48^\circ$.

\subsection{Latitude dependency of the angular momentum loss}

Once we know the local mass flux and the shape of the surface, it is possible to compute the local loss of angular momentum induced by the stellar winds, for a given angular velocity of the surface (let us recall that we suppose in this work that the angular velocity of the surface $\Omega$ is constant over the whole stellar surface). 

\begin{figure*}[htb!]
\begin{center}
\includegraphics[width=.45\textwidth]{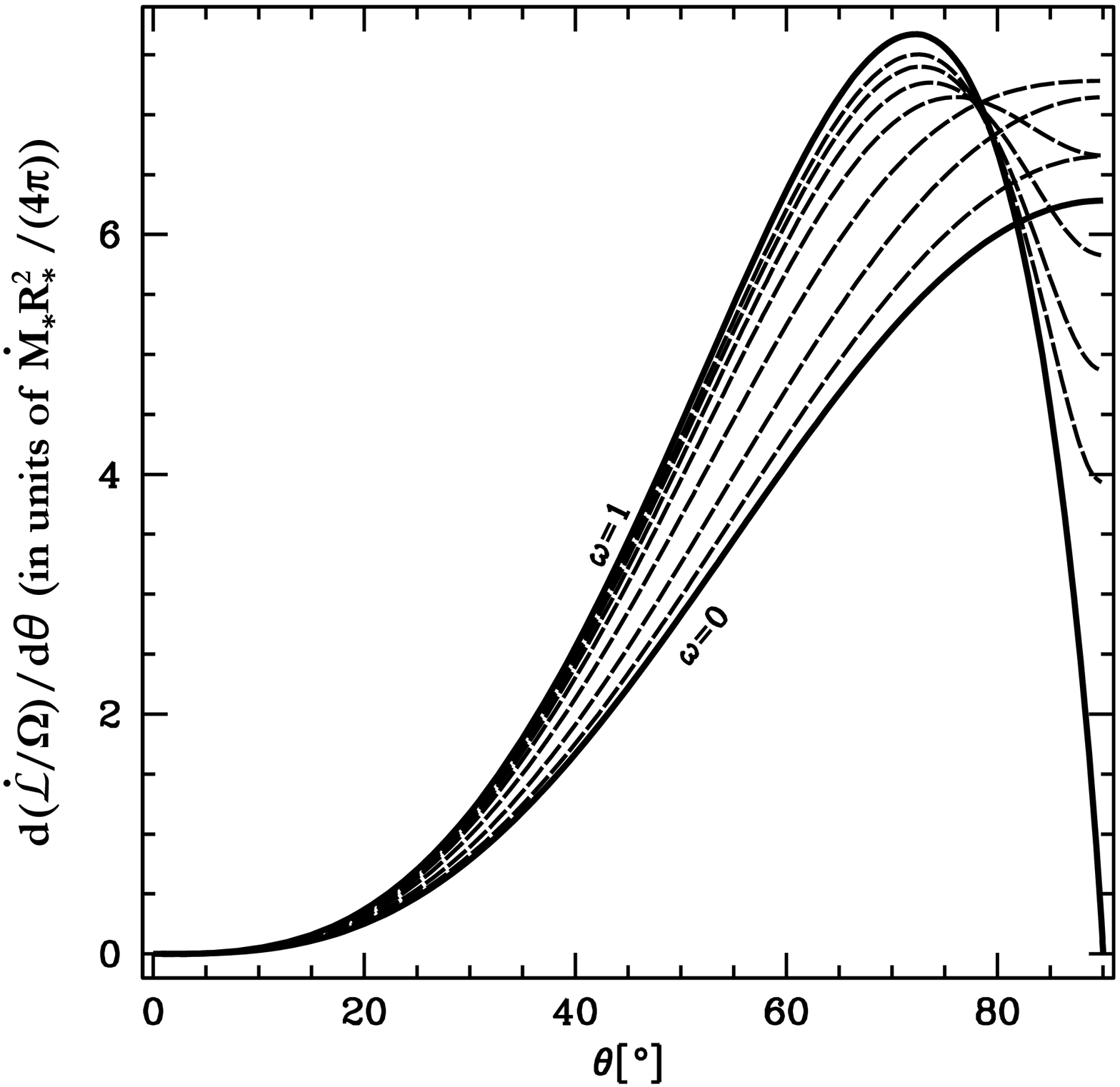}\hfill\includegraphics[width=.53\textwidth]{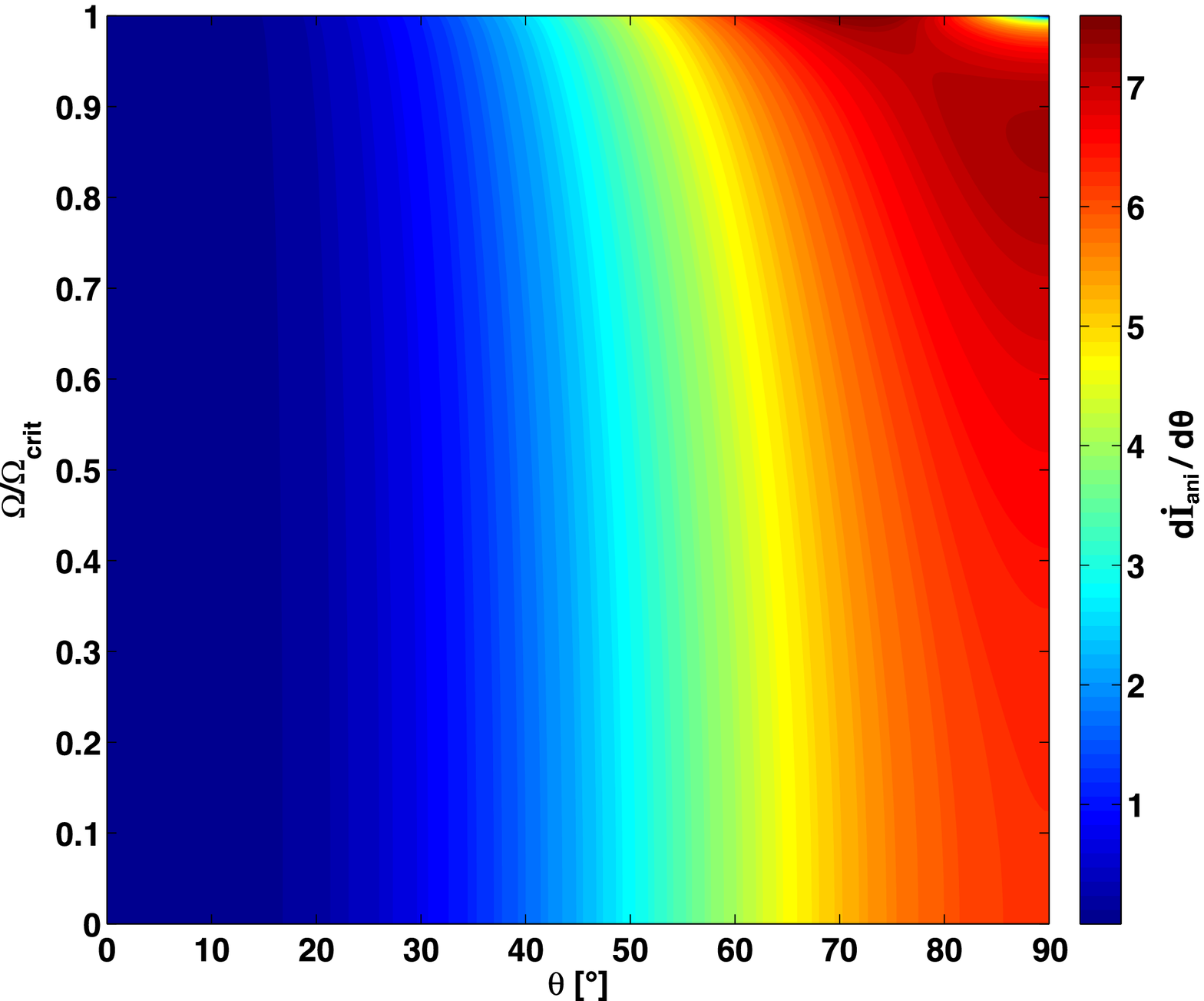}
\caption{\textit{Left panel: }Rate of angular momentum loss (normalised by the surface angular velocity) from infinitesimal rings centred on the rotational axis for each co-latitude $\theta$. The width of each ring is given by $r\mathrm{d}\theta$. The rotation parameter $\omega$ is indicated for the curves with $\omega = 0$ and $\omega = 1$. The intermediate dashed curves are for $\omega = 0.5,\,0.75,\,0.9,\,0.96,\,0.98,\,0.99,\,0.995$ respectively. The rate of angular momentum loss is expressed in terms of the normalised units $\dot{M}_\star R_\star^2 / (4\pi)$ where $\dot{M}_\star$ is the mass loss rate of the star, and $R_\star$ the polar radius of the star. \textit{Right panel: } Same as left panel in a 2D colour plot. The $x$-axis is the co-latitude, and the $y$-axis the ratio $\omega = \Omega/\Omega_\mathrm{crit}$. The colour scale on the right indicates the local angular momentum loss from infinitesimal rings centred on the rotational axis for each co-latitude $\theta$. The units are the same as in left panel. The smaller values are in blue and the larger ones in red.}
\label{IDist}
\end{center}
\end{figure*}

The loss of angular momentum per surface unit and time is given by:
\begin{equation}
\frac{\mathrm{d}\dot{\mathcal{L}}}{\mathrm{d}\sigma}=\frac{\Delta \dot{M}}{\Delta \sigma}(\theta)\Omega \mathcal{R}^2(\theta),
\end{equation}
where $\mathcal{R}(\theta)$ is the distance from the considered unit surface element to the rotation axis at the co-latitude $\theta$. Using the surface element (see above), and integrating over $\varphi$ to obtain only the co-latitudinal variation of the angular momentum loss, we have:
\begin{equation}
\frac{\mathrm{d}\dot{\mathcal{L}}}{\mathrm{d}\theta} = 2\pi\frac{\Delta \dot{M}}{\Delta \sigma}(\theta)r^4(\theta)\Omega\frac{\sin^3(\theta)}{\cos(\varepsilon)},\label{dLdthe}
\end{equation}
which is the contribution to the total angular momentum loss of an infinitesimal ring at the co-latitude $\theta$ and extending over an angle $\mathrm{d}\theta$. In order to avoid the $\Omega$-dependency, and as in our model, the surface of the star rotates at a constant angular velocity, we consider further the ratio $\dot{\mathcal{L}}/\Omega$. This permits to easily compare models with various rotation parameters. The distribution of the ratio $\dot{\mathcal{L}}/\Omega$ brought away by the wind is shown in Fig.~\ref{IDist}.

In this figure, we see, how the angular momentum flux (normalised by the surface angular velocity) is distributed as a function of the co-latitude $\theta$ for various rotation parameter $\omega$. Two effects are in concurrence: first, the increase of the equatorial radius (see Fig.\ref{FigSurf}), which increases the angular momentum flux near the equatorial regions, and second, the decrease of the local mass loss rate near the equator, which decreases the angular momentum flux in the same area.

Without rotation, the mass loss rate per surface unit is constant over the whole surface of the star, and the angle $\varepsilon$ between the effective gravity direction and the radial direction is zero. Examining eq.~(\ref{dLdthe}), we see that $\mathrm{d}\left(\frac{\dot{\mathcal{L}}}{\Omega}\right) / \mathrm{d}\theta$ varies as $\sin^3(\theta)$ (since all other terms are constant). The corresponding curve is labelled $\omega=0$ in Fig.~\ref{IDist}. Progressively increasing the rotation parameter, we see that the deformation of the stellar surface produces an increase of the angular momentum loss in the equatorial region. Once the rotation parameter $\omega \simeq 0.75$, the increase of the equatorial radius becomes counterbalanced by the progressive decrease of the local mass loss flux in the same region. The angular momentum loss becomes thus more and more reduced at the equator, and the maximum of momentum loss is shifted towards the pole, up to a co-latitude of $\sim 70^\circ$ when the rotation becomes critical. At that moment, no more angular momentum is lost at the equator, since there is no equatorial mass loss through radiative winds in this regime in our model.

\subsection{Total angular momentum lost}

Once the distribution of the angular momentum loss is known, we can integrate it over the co-latitude to obtain the total angular momentum loss rate due to the stellar winds:
\begin{equation}
\dot{\mathcal{L}}=2\pi\int_0^\pi \frac{\Delta \dot{M}}{\Delta \sigma}(\theta)r^4(\theta)\Omega\frac{\sin^3(\theta)}{\cos(\varepsilon)}\mathrm{d}\theta\label{Llossfull}.
\end{equation}
To well understand the effect of the anisotropic winds on the total angular momentum loss, we distinguish the following cases:
\begin{itemize}
\item \textbf{Case 1}:  we determine a mean stellar radius $r_\mathrm{mean}$ using the following relation:
\begin{equation}
L=\Sigma\sigma \left<T_\mathrm{eff}^4\right>\equiv 4\pi r_\mathrm{mean}^2\sigma \left<T_\mathrm{eff}^4\right>\label{rmean}
\end{equation}
where $L$ is the stellar luminosity and $\Sigma$ the total stellar surface. We neglect the stellar deformation and the wind anisotropy, and the loss of angular momentum is thus computed  on a sphere of radius $r_\mathrm{mean}$: $\dot{\mathcal{L}}=\frac{2}{3}\dot{M}\Omega r_\mathrm{mean}^2$. This radius is the one we would find if we measure the luminosity and the effective temperature of the star, and suppose that it is perfectly spherical. It gives the angular momentum loss as computed in numerical models where the effects of rotation on the shape of the surface and the mass loss distribution is neglected;
\item \textbf{Case 2}: the deformation of the star is accounted for, but the mass loss is uniformly distributed over the stellar surface (\textit{i.e.} the winds anisotropy is not taken into account). This case is academical, but is interesting in the sense that it allows to see the effect of the deformation only;
\item \textbf{Case 3}: both the deformation of the shape of the surface and the anisotropy are accounted for. The angular momentum loss is computed with relation \ref{Llossfull}.
\end{itemize}

\begin{figure}[t]
\begin{center}
\includegraphics[width=9cm]{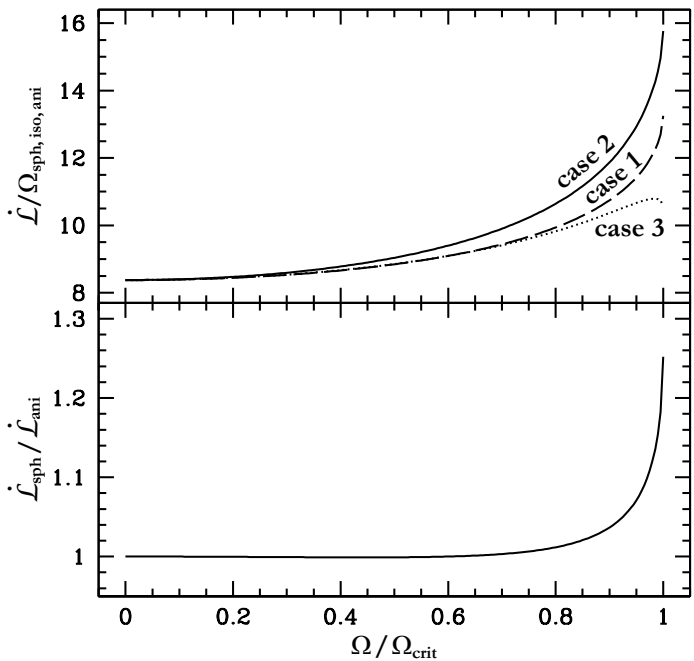}
\caption{\textit{Top panel: }Evolution of the total angular momentum loss as a function of the rotation rate. The long-dashed curve is the angular momentum loss when the mass loss is supposed isotropic and spherically symmetric ($\dot{\mathcal{L}}_{sph)}$, case 1). The solid curve is the angular momentum loss when the deformation of the star is accounted for, but the mass loss is assumed to be isotropic ($\dot{\mathcal{L}}_{iso}$, case 2). The dotted curve is the angular momentum loss when both the deformation and the anisotropy of the winds are accounted for ($\dot{\mathcal{L}}_{ani}$, case 3) (see text for more details). \textit{Bottom panel: }$\dot{\mathcal{L}}_{iso} / \dot{\mathcal{L}}_{ani}$ as a function of $\omega$.}
\label{FigInert}
\end{center}
\end{figure}

The results we obtain are shown in Fig.~\ref{FigInert}. The top panel shows the variation of the ratio $\mathcal{\dot{L}}/\Omega$ removed by the wind in the cases ``neither deformation, nor anisotropy", computed with the mean radius discussed above ($\mathcal{\dot{L}}/\Omega_\mathrm{sph}$, case 1),  ``deformation only" ($\mathcal{\dot{L}}/\Omega_\mathrm{iso}$, case 2), and ``deformation + wind anisotropy" ($\mathcal{\dot{L}}/\Omega_\mathrm{ani}$, case 3). In the three cases, the value for $\omega = 0$ is given by the integration of eq. \ref{Llossfull}, with $\Delta\dot{M}/\Delta\sigma = 1$, $r = 1$ and $\varepsilon = 0$ at each colatitude, leading to $\dot{\mathcal{L}}/\Omega = 2\pi \int_0^\pi \sin^3(\theta)\mathrm{d}\theta = 8\pi/3$.

The increase of $\mathcal{\dot{L}}/\Omega_\mathrm{sph}$ in case 1 is entirely due to the increase of the mean radius $r_\mathrm{mean}$ when $\omega$ increases. When the deformation is accounted for, we see that more angular momentum is lost because most of the mass leaves the surface of the star at a greater distance from the rotational axis. We see that, at the critical limit, deformation would increase the angular momentum loss rate by around $20\%$ with respect to case 1. For case 3, we see that the wind anisotropy largely compensates for the effect of the deformation and decreases the rate of angular momentum loss by $49\%$ with respect to case 2 and by $25\%$ with respect to case 1.

The lower panel of Fig.~\ref{FigInert} shows the ratio $\dot{\mathcal{L}}_\mathrm{sph} / \dot{\mathcal{L}}_\mathrm{ani}$. It shows the real impact of the account for the anisotropy of winds and the deformation of the stellar shape, compared with a model where we consider an isotropic spherical wind on the surface, with a radius determined by the stellar luminosity and mean effective temperature (as most of the stellar evolution codes).

Interestingly enough, the error committed on the angular momentum loss when neglecting the effects of wind anisotropies is small in most of the cases. It is less than $4\%$ if $\omega < 0.9$. At the critical velocity, the error is larger. Up to $25\%$ more angular momentum can be kept in the star when the effects of wind anisotropies are accounted for. Therefore, the effects of the wind anisotropies become important only for the faster rotators. This indicates that for most of the cases studied in stellar evolution, the precise account for the anisotropies are not relevant, and the errors induced by neglecting it will remain small.

\section{Effects of fast rotation on angular momentum loss: a numerical approach}\label{evolution}

The analytic relations above can provide some orders of magnitude estimates of the impact of the wind anisotropies on the loss of angular momentum assuming that $\omega$ remains more or less constant as a function of time.  To obtain more accurate values, it is necessary to compute numerical stellar models.

An additional complication, let aside in the above estimate, comes from the fact that the global mass loss rate increases with faster rotation. We first consider a case where the mass loss is kept constant. This will allow to make a more direct comparison with the semi-analytical results obtained above and thus to check that the implementation of the process was done correctly in the stellar evolution code. In a second step (see Sect.~\ref{MdotVink}), we shall consider the case of a model with all the usual prescriptions, in particular accounting for the evolution of the mass loss rate as a function of time, and the possible variation of the force multiplier parameters $\alpha$ and $k$ over the stellar surface.

\subsection{Models with constant mass loss rate}

We examine the effects of the anisotropic stellar winds on the evolution towards the critical velocity of two sets of models of $9\,\mathrm{M}_\odot$ star. To see the effects of anisotropic winds already on the zero age main sequence (ZAMS), we start the computation of the stellar models at a very high initial angular velocity of $80\%$ of the critical velocity. The metallicity is taken equal to $Z=0.002$, \textit{i.e.}, equivalent to that of the Small Magellanic Cloud (this implies smaller mass loss rates than at higher metallicity, favouring the reaching of the critical velocity for the numerical models presented in Sect.~\ref{MdotVink}). The rotation is treated as in \citet{Maeder2005a}, accounting for the internal magnetic field and its impact on the transport of angular momentum \citep{Spruit2002a}. These numerical models are based on the shellular rotation assumption \citep{Zahn1992a}. Even if the surface of our models reach rotation parameters close to $1$, as a large part of the stellar interior rotates far from the critical velocity, we consider that this assumption is valid. The account for the magnetic field ensures a strong coupling between the centre and the surface of the star, and leads to higher surface velocity. We expect thus a more important effect of the anisotropic winds in this context. Both models were computed using a  constant mass loss rate of $10^{-9}\,\mathrm{M}_\odot \mathrm{yr}^{-1}$, independent of the stellar surface parameters, and independent of the rotation rate. One  model was computed with the account for the anisotropic winds, and one assuming isotropic winds. In both cases, the deformation of the shape of the star is accounted for.

\begin{figure}[t]
\begin{center}
\includegraphics[width=9cm]{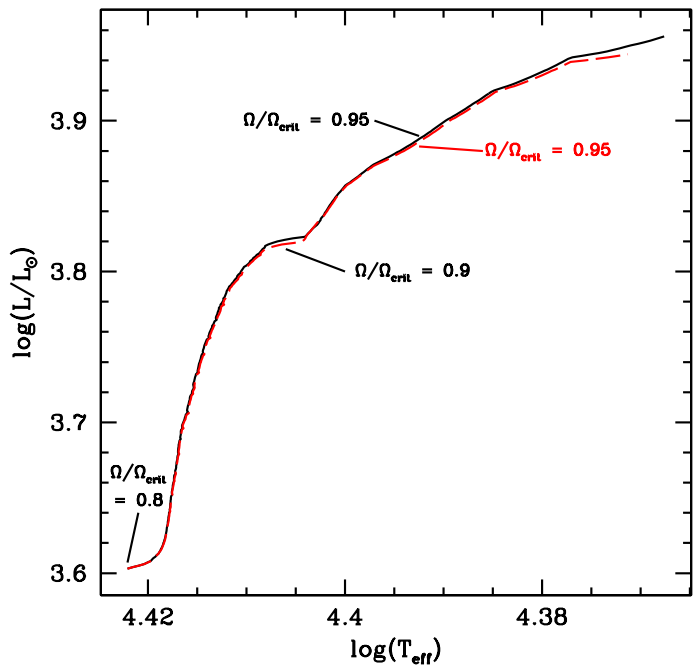}
\caption{HR diagram for the isotropic model (black solid line) and anisotropic model (red dashed line) with constant mass loss rate. The rotation parameter is indicated for some points along the tracks. The end point of the track (on the right) corresponds to the reaching of the first critical velocity.}
\label{HRD}
\end{center}
\end{figure}

\begin{figure}[t]
\begin{center}
\includegraphics[width=9cm]{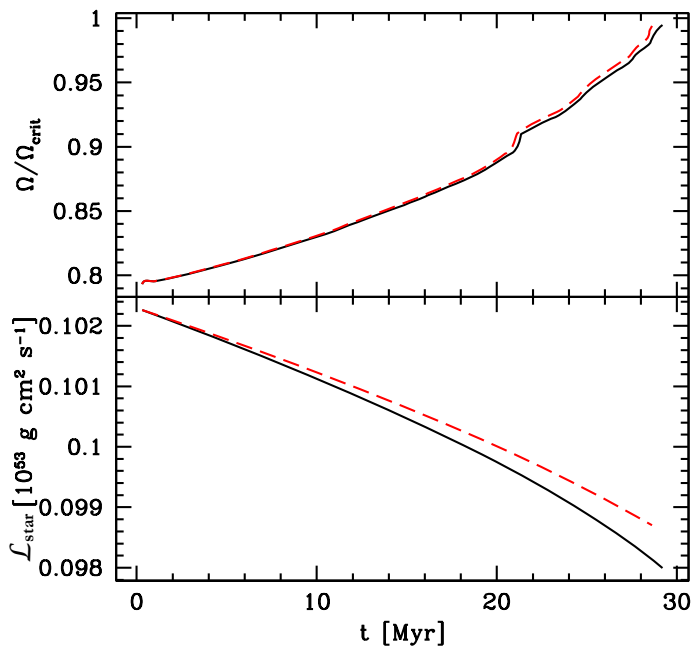}
\caption{\textit{Top panel: }Rotation parameter as a function of the time for the isotropic (black solid line) and anisotropic model (red dashed line) with constant mass loss rate. \textit{Bottom panel: } Total angular momentum contained in the star in unit of $10^{53}\,\mathrm{g}\,\mathrm{cm}^2\,\mathrm{s}^{-1}$. The models are represented as in top panel.}
\label{OmL}
\end{center}
\end{figure}

In Fig.~\ref{HRD}, we show the Hertzsprung-Russel diagram (HRD) of both models. The evolution stages where the models reach $80\%$, $90\%$ and $95\%$ of the critical angular velocity are indicated on the tracks. The ZAMS is bottom-left, and the evolution proceeds towards the top--right corner. We see that the account for the anisotropic winds has only a minor impact on the evolutionary track. A small deviation begins to appear when the surface velocity is around $90\%$ of the critical velocity. The anisotropic model evolves slightly more on the red side of the HRD. This is because this model rotates faster than the isotropic model (see the top panel of Fig.~\ref{OmL}). Its surface is slightly larger, and thus, for a given luminosity, the mean effective temperature will be lower.

Fig.~\ref{OmL} shows on the top panel the evolution of the ratio $\Omega/\Omega_\mathrm{crit}$ as a function of time for both models. The bottom panel shows the total angular momentum of the star. As expected, the mass loss due to the stellar winds causes a decrease of the angular momentum kept in the star. The model with the account for the wind anisotropy loses less angular momentum than the isotropic model. As a consequence, the stellar surface of the anisotropic model rotates slightly faster than the isotropic model.

The mean $\Omega/\Omega_\mathrm{crit}$ for the anisotropic model is $\overline{\Omega/\Omega_\mathrm{crit}} = 0.867$. According to Fig.~\ref{FigInert}, we expect that the anisotropic model keeps $1.12$ times more angular momentum than the isotropic one. During the time between the ZAMS and the reaching of the critical rotation rate, the anisotropic star loses an amount of angular momentum $\Delta\mathcal{L}_\mathrm{ani} = 3.56\cdot 10^{50} \,\mathrm{g}\,\mathrm{cm}^2\,\mathrm{s}^{-1}$. During the same time, the isotropic model loses $\Delta\mathcal{L}_\mathrm{iso} = 4.13\cdot 10^{50} \,\mathrm{g}\,\mathrm{cm}^2\,\mathrm{s}^{-1}$. The final ratio $\Delta\mathcal{L}_\mathrm{iso}/\Delta\mathcal{L}_\mathrm{ani} = 1.16$ is very close to the estimate based on the mean $\Omega/\Omega_\mathrm{crit}$.

\subsection{Total mass loss rate as a function of $\omega$ and $\Gamma_\mathrm{Edd}$}

\begin{figure}[t]
\begin{center}
\includegraphics[width=9cm]{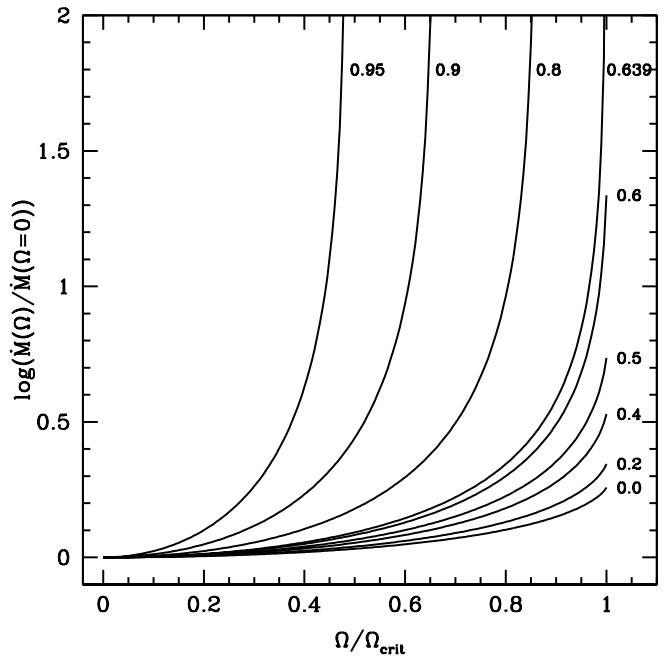}
\caption{Logarithm of the ratio $\dot{M}(\omega)/\dot{M}(\omega=0)$ as a function of $\omega = \frac{\Omega}{\Omega_\mathrm{crit}}$ for various values of the Eddington factor (value indicated at the top of each curve). For $\Gamma_\mathrm{Edd} \ge 0.639$ the curve tends towards infinity when $\omega$ approaches $\omega_\mathrm{max}$ (see Table 1). $\alpha$ is set to $0.43$ (see text).}
\label{MdotOmega}
\end{center}
\end{figure}

Before discussing the results of models accounting for time dependent mass loss rates, let us briefly recall how rotation enhances the global mass loss rate. Examining eq.~(\ref{MM0}), we see that the mass loss rate of a rotating star is simply expressed as a function of the angular velocity $\Omega$, the classical Eddington factor $\Gamma_\mathrm{Edd}$ and the local Eddington factor $\Gamma_\Omega$. Let us introduce in this relation the rotation parameter $\omega$ as defined above, and the definition of $\Gamma_\Omega$ given by relation (\ref{GammaOmega}):
\begin{equation}
\frac{\dot{M}(\Omega)}{\dot{M}(\Omega=0)} = \frac{\left(1-\Gamma_\mathrm{Edd}\right)^{\frac{1}{\alpha}-1}}{\left(1-\frac{4\omega^2V}{27\pi R_\mathrm{pb}^3}-\Gamma_\mathrm{Edd}\right)^{\frac{1}{\alpha}-1}}.\label{Mdotomega}
\end{equation}
To obtain Eq.~(\ref{Mdotomega}), we used eq~(\ref{vcrit1}), and replaced the mean density $\rho_\mathrm{m}$ by $V/M$, with $V$ the volume enclosed by the stellar surface.

For stars having an Eddington factor $\Gamma_\mathrm{Edd}$ larger than $0.639$, we can use  the expression of $v_\mathrm{crit,2}$ in Eq.~(\ref{vcrit2}) to rewrite the last relation:
\begin{equation}
\frac{\dot{M}(\Omega)}{\dot{M}(\Omega=0)} =\frac{\left(1-\Gamma_\mathrm{Edd}\right)^{\frac{1}{\alpha}-1}}{\left(\left(1-\Gamma_\mathrm{Edd}\right)\left(1-\frac{v_\mathrm{eq}^2}{v_\mathrm{crit,2}^2}\frac{V}{V_\mathrm{b}}\frac{R_\mathrm{eb}^2}{R_\mathrm{e}^2}\right)\right)^{\frac{1}{\alpha}-1}}.\label{Mdotomega2}
\end{equation}
Thus, we see that  the mass loss enhancement is governed by the ratio $v/v_\mathrm{crit,2}$, while the deformation of the star is governed by the ratio $v/v_\mathrm{crit,1}$. In the expression of $v_\mathrm{crit,2}$ intervenes the Eddington factor $\Gamma_\mathrm{Edd}$, thus the global enhancement factor of the mass loss will depend on two parameters $\omega$ (or $v_\mathrm{eq}/v_\mathrm{crit,2}$) and $\Gamma_\mathrm{Edd}$.

\begin{table}
\begin{center}
\caption{Maximum rotation parameter $\omega_\mathrm{max}$, and maximum increase of the mass loss rate as a function of the Eddington factor $\Gamma_\mathrm{Edd}$ (see text).}
\label{Mdotvalue}
\begin{tabular}{c|c|c||c|c|c}
$\Gamma_\mathrm{Edd}$ & $\omega_\mathrm{max}$ & $\dot{M}(\omega)/\dot{M}(0)$ & $\Gamma_\mathrm{Edd}$ & $\omega_\mathrm{max}$ & $\dot{M}(\omega)/\dot{M}(0)$\\
\hline
$0.0$ & $1.0$ & $1.810$ & $0.6$ & $1.0$ & $21.696$\\
$0.1$ & $1.0$ & $1.972$ & $0.639$ & $1.0$ & $\infty$\\
$0.2$ & $1.0$ & $2.214$ & $0.7$ & $0.968$ & $\infty$\\
$0.3$ & $1.0$ & $2.612$ & $0.8$ & $0.861$ & $\infty$\\
$0.4$ & $1.0$ & $3.383$ & $0.9$ & $0.659$ & $\infty$\\
$0.5$ & $1.0$ & $5.444$ & $0.95$ & $0.484$ & $\infty$\\
\hline
\end{tabular}
\end{center}
\end{table}

Fig.~\ref{MdotOmega} shows the variation of the ratio $\dot{M}(\omega)/\dot{M}(\omega=0)$ as a function of these two parameters. For this plot, we took a value for $\alpha=0.43$, which is adapted for effective temperatures $4.05 \le \log(T_\mathrm{eff}) \le 4.3$ (Lamers 2004, private communication). The volume $V$ enclosed by the stellar surface is numerically computed using eq.~(\ref{Shape}) for the shape of the surface

For Eddington factors greater than $0.639$, $v_\mathrm{crit,2} < v_\mathrm{crit,1}$ \citep{Maeder2000a}, thus the second limit is the one to be considered. At this limit, the mass loss rate becomes very high. The precise value of the enhancement cannot be given since, at that limit and beyond, the hypothesis made in deriving eq.~(\ref{Mdotomega}) no long holds. For instance, near the Eddington limit, continuous radiation field contributes in pushing out the outer layers, while for obtaining the above expression, we made the hypothesis that the wind is triggered by radiation pressure on lines.

In Table~\ref{Mdotvalue}, we indicate the maximum values of the rotation parameter $\omega_\mathrm{max}$ above which the formulae given before for the enhancement of the mass loss rate due to rotation do no long hold. For values of $\Gamma_\mathrm{Edd}$ inferior to $0.639$, the maximum value is equal to one.  In that range, stars with $\omega$ reaching  $1$ will begin to lose mass through a mechanical mass loss in the equatorial regions. For $\Gamma_\mathrm{Edd}$ superior to $0.639$, the maximum value is inferior to one. In that domain, the continuous emission will participate in pushing out the matter and very important mass loss rates are expected \citep{vanMarle2008b}. Depending on the value of $\Gamma_\mathrm{Edd}$, the winds can be more or less anisotropic: for $\Gamma_\mathrm{Edd}$ values just above $0.639$, $v_\mathrm{crit,2}$ is near $v_\mathrm{crit,1}$ and strong anisotropies are expected; when $\Gamma_\mathrm{Edd}$ is near $1$, $v_\mathrm{crit,2}$ is much lower than $v_\mathrm{crit,1}$ and the winds are expected to be isotropic.

\subsection{Models with realistic mass loss rate}\label{MdotVink}

\begin{figure}[t]
\begin{center}
\includegraphics[width=9cm]{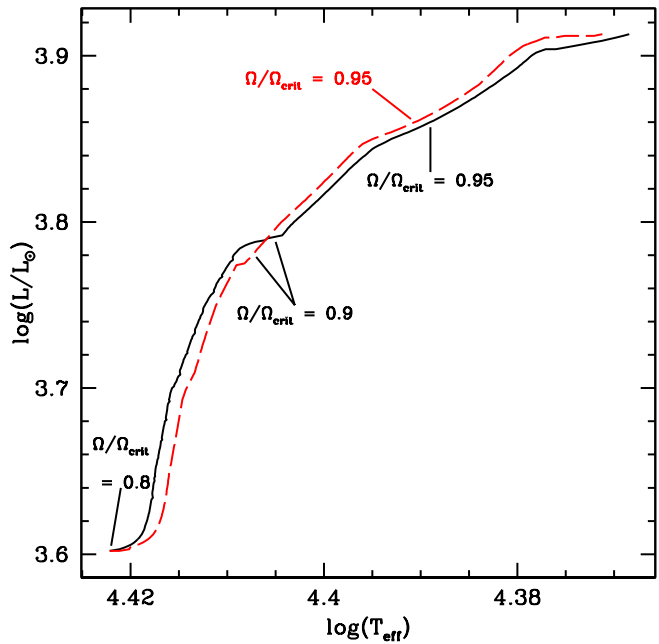}
\caption{HR diagram for the isotropic model (black solid line) and anisotropic model (red dashed line) with realistic mass loss rate. The rotation rate is indicated for some points along the tracks.}
\label{HRDreal}
\end{center}
\end{figure}

\begin{figure}[t]
\begin{center}
\includegraphics[width=9cm]{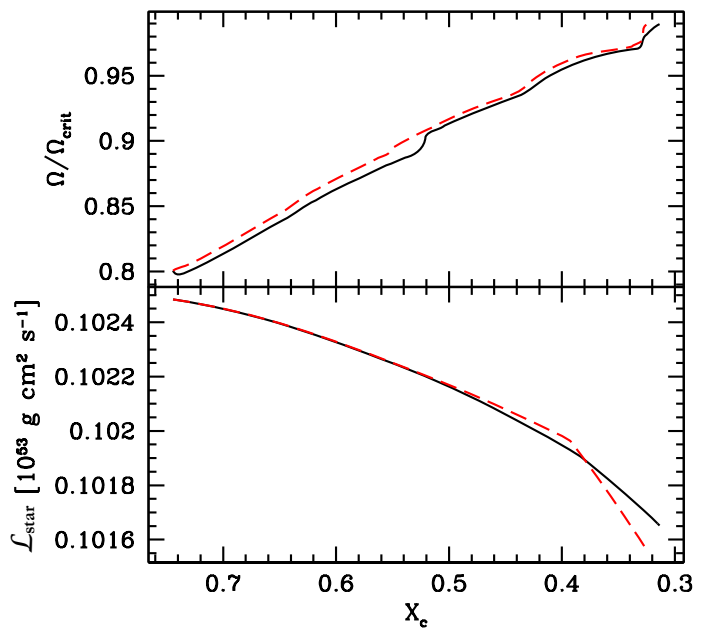}
\caption{\textit{Top panel: }Rotation rate as a function of the central hydrogen mass fraction for the isotropic (black solid line) and anisotropic model (red dashed line) with realistic mass loss rate. The evolution proceeds from left to right. \textit{Bottom panel: } Total angular momentum contained in the star in unit of $10^{53}\,\mathrm{g}\,\mathrm{cm}^2\,\mathrm{s}^{-1}$. The models are represented as in top panel.}
\label{OmLreal}
\end{center}
\end{figure}

Here we discuss $9\,\mathrm{M}_\odot$ models using a realistic mass loss rate \citep{Vink2001a}, and accounting for the increase of the mass loss rate induced by rotation (see above). One model takes the effect of anisotropic winds into account, and one has an isotropic mass loss. We also included in the anisotropic model the variation of the force multiplier parameters $k$ and $\alpha$ due to the variation of the local effective temperature as a function of the co-latitude. Models were  followed until they reached the critical velocity.

The variations of the force multiplier parameters are accounted for in the following way. In the expression of the increase of the global mass loss rate due to rotation (\ref{Mdotomega}), we take $\alpha$ and $k$ given by the mean effective temperature of the star. For the computation of the anisotropic effects (\ref{Mdotsurf}), we use at each co-latitude $\alpha$ and $k$ corresponding to the local effective temperature, allowing variations over the stellar surface.

Fig.~\ref{HRDreal} shows the HRD for this set of models. The black curve represents the model with an isotropic mass loss over the surface, and the red curve the anisotropic one. As in the previous case, the tracks in this diagram are very similar for both models, even if more complete physics is included. This confirms that even with full treatment of the wind anisotropy, including a realistic mass loss rate and the variation of the force multiplier parameters, the effect of the anisotropic mass loss for very fast rotators remains very small.

On the top panel of Fig.~\ref{OmLreal}, we see the evolution of the ratio $\omega$ as a function of the central hydrogen content for the isotropic model (black solid line) and the anisotropic one (red dashed line). The ZAMS is on the left, and the point where $\omega = 1$ is on the right. The anisotropic model rotates ever faster than the isotropic one, but the difference is limited. The increase of rotation of the isotropic model near $X_\mathrm{c} = 0.52$ causes the crossing of the tracks in the HRD (see Fig.~\ref{HRDreal}). The inflation of the surface induced by the higher rotation rate decreases the mean effective temperature, pushing the track on the right.

The bottom panel of Fig.~\ref{OmLreal} shows an interesting feature. From the ZAMS until $X_\mathrm{c} \sim 0.4$, the behaviour of the total angular momentum contained in the whole star is similar to the models with constant mass loss rate: the anisotropic model keeps more angular momentum due to the wind being polar. From that point on, however, the star reaches a high enough rotation rate to strongly decrease the equatorial effective temperature. The force multiplier parameters are different in this area, and generate a strong equatorial mass loss. This effect produces the change in the angular momentum loss rate: the anisotropic model loses more angular momentum than the isotropic model. When the star reaches the critical limit, the isotropic model finally has a higher angular momentum content than the anisotropic model!

The final angular momentum of the star strongly depends on the angular momentum removed by the mechanical mass loss that the star undergoes during the critically rotating phase. It is difficult to estimate which of the isotropic or anisotropic model will have the higher content at the end of the stellar evolution, and to quantify this difference without a model accounting for the mechanical mass loss. This question, and first estimates of the mass lost in the equatorial disk, will be addressed in a forthcoming paper.

\section{Conclusion}\label{conclusion}

The main result of this paper is that radiative wind anisotropies do not strongly affect the angular momentum content of stars, in contrast with previous estimates. The different conclusion obtained here comes mainly from a precise account of the effect of the surface deformation in addition to the effects induced by  the variation of the mass flux with the co-latitude. Interestingly, taking into account the variation of the force multiplier parameters over the surface when the star is near the critical limit can favour an equatorial-enhanced mass loss rather than a polar mass loss. In that case, the angular momentum loss when the effects of wind anisotropies are accounted for can be higher than when they are neglected!

Since the anisotropic winds do have a strong influence on the evolution of the star, the strong enhancement of the polar mass flux has a big effect on the evolution of the circumstellar medium \citep[see][]{Georgy2009a}. The formation of an asymmetric nebula around fast rotating stars is likely.

However, another point which appeared in that work is the importance of the equatorial mass loss triggered by the reaching of the first critical limit. Two effects can act to keep the star at the critical limit: the first is the variation of the force multiplier parameters $\alpha$ and $k$ in the equatorial regions when the local effective temperature becomes low enough due to the effect of rotation. This triggers strong equatorial radiative winds, and is already accounted for in this study. The other is the mechanical mass loss in the equatorial plane, when the equatorial effective gravity vanishes. In that case, the mechanical mass loss through an equatorial disc can remove angular momentum. In a forthcoming paper we shall study in a quantitative way the impact of such a disc mass loss on the evolution of fast rotating stars.

\bibliographystyle{aa}
\bibliography{MyBiblio}

\end{document}